\documentclass[preprint]{revtex4}
\nonstopmode

\usepackage{amsmath}
\usepackage{amsbsy}
\usepackage{pstricks,graphicx}
\newcommand{\ket}[1]{\ensuremath{|#1\rangle}}
\newcommand{\bracket}[2]{\ensuremath{\langle #1|#2\rangle}} 
\newcommand{\dirint}[3]{\ensuremath{\langle #1|#2|#3\rangle}}

\newcommand{\bs}{\boldsymbol}

\begin{document}

\title{Modifying the
variational principle in the action integral functional
derivation of time-dependent density functional theory}
\author{J. Schirmer}
\affiliation{Theoretische Chemie,\\ Physikalisch-Chemisches Institut,
Universit\"{a}t  Heidelberg,\\
D-69120 Heidelberg, Germany}
\date{October 1, 2010}

\begin{abstract}
According to a recent paper by G. Vignale [Phys. Rev. A \textbf{77}, 062511 (2008)], 
the problems arising in the original derivation of time-dependent density
functional theory (TDDFT) based on the Runge-Gross (RG) action-integral functional (AIF) are 
due to an incorrect variational principle (VP). This argument and the proposed 
modification of the VP are critically analyzed. The more fundamental
problem, though, is the indefiniteness of the RG AIF. In contrast to a widely 
held belief, that indefiniteness is not eliminated in the variational procedure,
which unwittingly is corroborated by Vignale's initial point.
\end{abstract}

\maketitle

\section{Introduction}

The original foundation of time-dependent density-functional theory (TDDFT)
by Runge and Gross (RG)~\cite{run84:997} was based on a variational principle for 
the so-called action integral functional (AIF). It was later found that 
this route to TDDFT leads to a violation of causality in the kernel 
of the exchange correlation functional (``causality paradox'')~\cite
{Nalewajski:1996,raj96:3916,lee98:1280}
indicating a serious problem with the RG AIF. Van Leeuwen~\cite{lee01:1969} located the
problem in the arbitrary purely time-dependent (td) phase factor 
in the wave function that is not determined by 
the mapping between td densities and wave functions according to the first
Runge-Gross theorem (RG1). 
As a consequence, the RG AIF is ill-defined (see also Ref.~\cite{sch07:022513}).

While apparently van Leeuwen's analysis was not accepted unanimously in the TDDFT
community, there was a broad consensus that any flaws in the original foundation
would not affect the essence of TDDFT, because the td Kohn-Sham equations
at the core of the theory could be established directly via the RG1 
mapping theorem, evading the need for a variational principle. 
For a discussion of the problems arising in the latter mapping-based
foundation the reader is referred to Ref.~\cite{sch07:022513,sch08:056502}.

In a recent paper~\cite{vin08:062511}, henceforth referred to as paper I, G. Vignale
has attempted to rehabilitate the RG AIF and the TDDFT foundation associated with it.
Guided by a correct finding, he concluded that the 
original formulation 
of the variational principle is wrong and must be modified appropriately. 
In the following brief report we will analyze Vignale's work in some detail.
As will be seen, even when the problem of the definiteness of 
the RG AIF is ignored, the diagnosis in I is unfounded and the conclusions 
derived thereof not compelling.

\section{Definition of the action integral functional}

Let us first address the problem of the indefiniteness of the 
RG AIF reading  
\begin{equation}
\label{eq:actint}
A[n] = \int_{t_1}^{t_2} dt\, \dirint{\Psi[n](t)}{i\frac{\partial}{\partial t}
 - \hat{H}(t)}{\Psi[n](t)}
\end{equation}
Here $\Psi[n](t)$ is the wave function associated with the 
time-dependent (td) density $n(\bs{r},t)$ according to the  
RG1 theorem. It should be noted 
that the RG1 mapping depends also on the initial value of the
wave function, $\Psi(t_1)$; as usual, the initial state dependence
will not explicitly be indicated for notational brevity.

As correctly stated by RG, the wave function $\Psi[n](t)$
is determined by the density only up to a purely time-dependent phase factor $e^{i \alpha(t)}$.
However, the obvious consequence of this indefiniteness for the AIF 
has been falsely addressed in Ref.~\cite{run84:997} and disregarded in later work 
(see, for example, Gross and Kohn~\cite{gro90:255}).
Because of the time derivative in the integrand of the AIF,
there arises a constant additive term, 
$c = - \int_{t_1}^{t_2} \dot \alpha(t) dt = \alpha(t_1) - \alpha(t_2)$.
As the phase function $\alpha(t)$ is completely arbitrary so is the constant c. 
This means that the AIF, being a mapping of time-dependent densities (or density trajectories)
onto mere numbers, is ill-defined.

Unfortunately, this finding is often dismissed by stating that the additive constant 
is "immaterial" with regard to the variational condition (see Casida~\cite{cas95:155}).
A similar remark is in the  
brief Sec. IV of paper I: "While the absolute numerical value of the RG action has no 
physical meaning (because a multiplication of the wave function by an 
arbitrary phase factor changes its value by an arbitrary amount), it must be 
borne in mind that the action determines the dynamics through its 
variations, and those variations are independent of the arbitrary additive 
constant." It is noted (in parantheses) that "a 
similar situation occurs in classical mechanics, since the Lagrangian is 
defined up to an arbitrary total derivative with respect to time." 

However, the latter remark is not quite accurate. The variational principle in classical 
mechanics is based on the action integral 
\begin{equation}
\label{eq:classmech}
S = \int_{t_1}^{t_2} dt\, L(q(t),\dot q(t),t) \
\end{equation}
assuming for simplicity only one degree-of-freedom.
The requirement $\delta S = 0$ for variations of the coordinate function $\delta q(t)$ 
with $\delta q(t_1) = \delta q(t_2) = 0$ allows one to derive the well-known
(Euler) equation-of-motion (EOM)
\begin{equation}
\label{eq:Euler}
\frac{d}{dt} \frac{\partial L}{\partial \dot q}  -  
\frac{\partial L}{\partial q} = 0
\end{equation}
The correct statement with regard to the non-uniqueness of the Lagrangian reads
that $L$ is defined up to the total time derivative of an 
arbitrary function of the coordinates and time, say 
$\frac{d}{dt} f(q(t),t)$.  
The time 
integration leads to an additive constant $f(q(t_2),t_2) - f(q(t_1), t_1)$, for which 
the variation with respect to $\delta q(t)$ vanishes due to the requirement that 
$\delta q(t_1) = \delta q(t_2) = 0$, so that the result of the variational
procedure is not affected. Obviously, the non-uniqueness associated with terms of the type 
$\frac{d}{dt} f(q(t),t)$ is quite specific and differs from the 
vague formulation in paper I. As one can readily see, an arbitrary total derivative
with respect to time leads to 
\begin{equation}  
\label{eq:modvp}
\delta S = \int_{t_1}^{t_2} dt\,(\frac{\partial L}{\partial q} -
\frac{d}{dt} \frac{\partial L}{\partial \dot q}) \delta q(t)  + C = 0 
\end{equation}
Obviously, the arbitrary constant $C$ does not allow one to derive a
meaningful EOM.

In the case of the RG AIF the argument mentioned above insinuates that
$\delta c = 0$. As will be discussed at some detail below,
the main point made in paper I is that the variation
of the wave function, $\delta \Psi(t)$, associated with the variation of the underlying density,
$n(t)+ \delta n(t)$
(where $\delta n(t_1) = \delta n(t_2) = 0$),
vanishes for the initial time $t_1$ but not for the final time $t_2$. 
This means, in particular, that the variation of the purely td phase will
vanish at $t_1$ but not at $t_2$, that is, $\delta \alpha(t_2) \neq 0$. 
Obviously, the problem of the indefinite constant does not disappear in 
the variational procedure, and one cannot escape the 
conclusion that the RG AIF 
is ill-defined due to the ambiguity with respect to 
a purely time-dependent phase factor.  
It might be tempting to
dismiss that innocuous phase factor in the wave function 
as somehow irrelevant, but one should recall that 
the time-dependence of stationary states consists exactly in such phase 
factors. They are an indispensable ingredient in the solutions of the td 
Schr\"{o}dinger equation (SE).

Nothing meaningful can be deduced if the origin, that is, the
RG AIF is ill-defined, and we could stop at this point. 
Nevertheless it may be of interest to elaborate on Vignale's attempt 
to modify the VP based on the AIF without dismissing the latter from the outset.
So, tentatively, we will assume in the following that the RG AIF 
is a well-defined entity. Of course, the existence (definiteness) of 
the RG AIF is a concrete mathematical assertion being either true or false.
A way to prove a premise false is to inspect its implications to be deduced in 
a logically stringent way. In this respect it is important to analyze 
arguments in paper I.

\section{Variational principle for time-dependent quantum mechanics}

For a better understanding of the argumentation in paper I and
clarifying certain misconceptions arising there,  
a brief review of the td variational principles (VP) 
in quantum mechanics will be given here.
Of course, all of the following is well-known  
and a reader familiar with
the topic might safely skip this section.

The time-evolution of a wave function is determined by the td SE, 
together with an initial-state condition 
specifying the wave function at a given time. Often the exact solution of the
td SE is not feasible, and one has to resort to approximations, say, to a parametrized
form of the wave function, in which the parameters are time-dependent.
The question is how to obtain a suitable EOM in such a situation.
For this purpose one may resort to the well-known VPs of
td quantum mechanics~\cite{dir30:376,Frenkel:1934,mcl64:39,Kramer:1981}. In the so-called
Lagrangian form (see Ref.~\cite{Kramer:1981}), the VP is based on the 
action-type integral
\begin{equation}
\label{eq:vp1}
A[\Psi] = \int_{t_1}^{t_2} dt \dirint{\Psi(t)}{ i \frac{\partial}{\partial t}-\hat{H}}{\Psi(t)}   
\end{equation}
for the time-development of a wave function in a given (though arbitrary) time interval $[t_1,t_2]$. 
The aim is to determine an ``optimal'' wave function $\Psi (t)$ 
(more accurately, derive a pertinent EOM) such that
the action integral is stationary with respect to (small) 
variations of the form
$\tilde{\Psi}(t) = \Psi (t) + \delta \Psi (t)$. 
Here the variations are required to vanish at the boundaries of the time interval,
that is, $\delta \Psi (t_1) = \delta \Psi (t_2) = 0$. The VP
then reads 
\begin{equation}
\label{eq:vp2}
\delta A[\Psi] = 0, \,\text{where}\,\, \delta \Psi (t_1) = \delta \Psi (t_2) = 0
\end{equation}
The integrand in the action integral~(\ref{eq:vp1}) can be interpreted as 
the expectation value of the deviation with respect to the exact td SE. The obvious idea
underlying the VP~(\ref{eq:vp2}) is that the
desired wave function $\Psi(t)$ should minimize (make stationary) this deviation within
the given time interval.

A more convenient, namely, instantaneous form of the td VP may be obtained 
by rewriting $\delta A[\Psi]$ in such a way
that the VP applies directly to the integrand in the time-integration 
rather than to the action integral. Starting from 
\begin{equation}
\label{eq:vp3}
\delta A[\Psi] = \int_{t_1}^{t_2} dt\, 
\dirint{\delta \Psi(t)}{i \frac{\partial}{\partial t}-\hat{H}}{\Psi(t)} 
               +  \int_{t_1}^{t_2} dt\,
\dirint{\Psi(t)}{ i \frac{\partial}{\partial t}-\hat{H}}{\delta \Psi(t)} 
\end{equation}
one may replace the term involving $\delta \dot \Psi(t)$ using integration by parts,
\begin{equation}
\bracket{\Psi(t)}{\delta \dot \Psi(t)} = \frac{d}{dt}\bracket{\Psi(t)}{\delta \Psi(t)}
               - \bracket{\dot \Psi(t)}{\delta \Psi(t)}
\end{equation}
Because $\delta \Psi (t_1) = \delta \Psi (t_2) = 0$ the 
integral of the total time derivative vanishes, and one arrives at
\begin{equation}
\label{eq:vp4}
\delta A[\Psi] =  2 \int_{t_1}^{t_2} dt\, Re \,
\dirint{\delta \Psi(t)}{ i \frac{\partial}{\partial t}-\hat{H}}{\Psi(t)} = 0 
\end{equation}
Taking into account that the integration boundaries $t_1, t_2$ are arbitrary,
one may conclude (see Kucar \textit{et al.}~\cite{kuc87:525}) that 
the integral~(\ref{eq:vp4}) vanishes if and only if
\begin{equation}
 \label{eq:vp5}
 Re \, \dirint{\delta \Psi(t)}{ i \frac{\partial}{\partial t}-\hat{H}}{\Psi(t)} = 0 
\end{equation}
This establishes an instantaneous form of a td VP being fully equivalent to the
Lagrangian form according to Eqs.~(\ref{eq:vp1},\ref{eq:vp2}).
If together with the variations $\delta \Psi (t)$ also
the variations $i \delta \Psi (t)$ are permitted, one arrives at
the related form
\begin{equation}
\label{eq:vp6}
 Im \, \dirint{\delta \Psi(t)}{ i \frac{\partial}{\partial t}-\hat{H}}{\Psi(t)} = 0 
\end{equation}
Both forms are comprised in the well-known Dirac-Frenkel 
VP~\cite{dir30:376,Frenkel:1934}, reading
\begin{equation}
\dirint{\delta \Psi}{i \frac{\partial}{\partial t}-\hat{H}}{\Psi} = 0  
\end{equation}
This shows that the Lagrangian and the Dirac-Frenkel variational principles
are equivalent provided that the both $\delta \Psi (t)$ and $i\delta \Psi (t)$
are allowed variations of the wave function. For an illustrative application of
the Dirac-Frenkel VP the reader is referred to the derivation of the 
multi-configuration time-dependent Hartree
(MCTDH) equations by Beck \textit{et al.}~\cite{bec00:1}.

Obviously, an exact solution of the td SE fulfills the Dirac-Frenkel (and the Lagrangian)
VP. The converse proposition, however, is not true:  
In general, a wave function fulfilling the td VP need not be the exact solution 
of the td SE. As mentioned above, the purpose of the td VPs is to 
generate an EOM for cases where one cannot apply the td SE. 

\section{Density-based time-dependent variational principle}

Supposing that the 
RG AIF~(\ref{eq:actint}) is well-defined, we will now briefly discuss how,
at least in principle, a valid EOM for an optimal (or exact) 
td density would emerge from the pertinent VP, 
\begin{equation}
\label{eq:vpn1}
\delta A[n] = \int_{t_1}^{t_2} dt \, \dirint{\delta \Psi[n, \delta n](t)}
{ i \frac{\partial}{\partial t}-\hat{H}}{\Psi [n](t)}  
+ \int_{t_1}^{t_2} dt\, \dirint{\Psi[n](t)}
{ i \frac{\partial}{\partial t}-\hat{H}}{\delta \Psi [n, \delta n](t)}  = 0
\end{equation}
Here the wave function $\Psi[n](t)$ is assumed to be somehow parametrized
in terms of an underlying td density function $n(t)$.
Once again, the intuitive idea of the VP is to minimize (make stationary)
the expectation value of the td SE difference operator in a given time interval
with respect to the manifold of permitted td densities (density trajectories).
The optimal td density $n(t)$ is to be compared to 
deviating density trajectories  $\tilde{n}(t) = n(t) + \delta n(t)$, 
where $\delta n(t_1) = \delta n(t_2) = 0$.
The variations of the td density induces associated variations of the wave function,
reading (by definition)
\begin{equation}
\delta \Psi[n, \delta n](t) \equiv \Psi [n + \delta n](t) - \Psi [n](t)
\end{equation} 
Note that $\delta \Psi[n, \delta n](t)$ will depend both on $n(t)$ and $\delta n(t)$.
While $\delta \Psi[n, \delta n](t)$ vanishes at the initial time $t_1$,
this is not necessarily so at the final time, $t_2$, as Vignale rightly observes.
We will come back to this point below.

Now the question is how to proceed from here to an at least formally
compelling EOM for the optimal td density. 
For this purpose 
one might consider the possibility that $\delta A[n]$ can be written
in the form of a td functional derivative with respect to the density at
time $t$,
\begin{equation}
\label{eq:vpn2}
\delta A[n] = \int d\bs{r} \int_{t_1}^{t_2} dt f[n](\bs{r},t) \delta n(\bs{r},t) = 0
\end{equation}
where 
\begin{equation}
\label{eq:vpn3x}
 f[n](\bs{r},t) = \frac{\delta A[n]}{\delta n(\bs{r},t)}
\end{equation}
This would lead directly to the simple equation~\cite{run84:997}
\begin{equation}
\label{eq:vpn3}
\frac{\delta A[n]}{\delta n(\bs{r},t)} = 0
\end{equation}
However, such an outcome certainly is too simplistic. Even if one assumes 
that the wave function at a given time $t$ is determined by the density $n(t)$ at that
time, the first time derivative $\dot \Psi(t)$ and thus also $\dot n(t)$
is needed to evaluate the expectation value of the td SE difference operator at time $t$.
Therefore one has to expect that also a functional derivative with respect to 
$\dot n(t)$ will come into play leading to an expression of the type
\begin{equation}
\delta A[n] = \int d\bs{r} \int_{t_1}^{t_2} dt (f[n,\dot n](\bs{r},t) \delta n(\bs{r},t) 
+ g[n, \dot n](\bs{r},t) \delta \dot n(\bs{r},t)) = 0
\end{equation}
where $f[n,\dot n](\bs{r},t) = \frac{\delta A[n]}{\delta n(\bs{r},t)}$
and $g[n,\dot n](\bs{r},t) = \frac{\delta A[n]}{\delta \dot n(\bs{r},t)}$ are 
functional derivatives
with respect to $n(t)$ and $\dot n(t)$, respectively.
Using integration by parts (as in the derivation of the classical Euler equations)
and $\delta n(t_1) = \delta n(t_2) = 0$ it follows that
\begin{equation}
\delta A[n] = \int d\bs{r}\int_{t_1}^{t_2} dt \left ( f[n,\dot n](\bs{r},t)  
- \frac{d}{dt}g[n, \dot n](\bs{r},t) \right ) \delta n(\bs{r},t) = 0
\end{equation}
yielding the Euler-type equation 
\begin{equation}
 f[n,\dot n](\bs{r},t)  - \frac{d}{dt}g[n, \dot n](\bs{r},t) = 0
\end{equation}
In fact, the latter equation might be viewed as 
a viable EOM. Its practical use, however, would require a guess for an
AIF depending not only on $n$ but also on $\dot n$.

Still, this may not yet be the end of the line. As was discussed in Ref.~\cite{sch07:022513},
one may need $\dot n(t)$ to determine $\Psi(t)$, so that the evaluation of the 
AIF integrand at a given time $t$ would even require
the second time derivative $\ddot n(t)$.
The corresponding expansion of $\delta A[n]$ would read
\begin{equation}
\delta A[n] = \int d\bs{r}\int_{t_1}^{t_2} dt 
 \left [ \left ( f[n,\dot n, \ddot n](\bs{r},t)  
- \frac{d}{dt}g[n, \dot n, \ddot n](\bs{r},t)\right ) \delta n(\bs{r},t) + 
h[n, \dot n, \ddot n](\bs{r},t) \delta \ddot n(\bs{r},t)\right ] = 0
\end{equation}
where $h[n, \dot n, \ddot n](\bs{r},t) = \frac{\delta A[n]}{\delta \ddot n(\bs{r},t)} $.
In the latter case it is not clear how a viable EOM could be deduced at all.

To conclude, even if the RG AIF were well-defined it would not be guaranteed that
a valid EOM for the td density can be deduced. Wether this is possible or not, 
depends crucially on a further specification of the RG mapping 
$n(t) \rightarrow \Psi[n](t)$. What pieces of information are needed to generate
the wave function at a given time $t$ from the density trajectory? 
Does $n(t)$ suffice, are both $n(t)$ and $\dot n(t)$ needed, or is the situation even more complex,
requiring, e.g., the past or even the future of the density trajectory with regard to $t$?
So far, these questions have never been properly addressed let alone clarified.

\section{Modification of the variational principle?}

After the preceding considerations we are now prepared
to review the proposal for a modification of the AIF VP as presented in the 
basic Sec. II of paper I. Equations in I will be indicated here by
adding the prefix I- to the original number.

In the beginning the td VP in the Lagrangian form~(\ref{eq:vp1})
is considered, implying that this VP is
equivalent to the td SE. As we have argued above
the latter statement is misleading. Of course, the VP
is based on the td SE in that the expectation value of the
td SE difference operator is to be minimized (in a given time interval).
As a result, a solution of the td SE fulfills the td VP,
but the converse is not true: a wave function fulfilling the td VP
must not necessarily be a solution of the td SE. 
The ``proof'' given in I (second column of p. 2) for that
incorrect assertion
proceeds in the familiar way used to demonstrate the equivalence of
the Lagrangian and Dirac-Frenkel VPs (see above), but ends after the 
time integration by parts with the following equation (I-6):
\begin{equation}
\label{eq:I6}
\delta A[\Psi] =   \int_{t_1}^{t_2} dt\,
\dirint{\delta \Psi(t)}{ i \frac{\partial}{\partial t}-\hat{H}}{\Psi(t)}
+ \int_{t_1}^{t_2} dt \,\bracket{(i \frac{\partial}{\partial t}-\hat{H}) \Psi(t)}{\delta \Psi(t)}
+ i \bracket{\Psi(t)}{\delta \Psi (t)}\mid ^{t_2}_{t_1} = 0 
\end{equation}
Note that here we have adopted slight notational changes (e.g. using the times $t_1,t_2$ rather than
$0, T$). Actually, the last term on the r.h.s. vanishes because $\delta \Psi (t)$ vanishes at
the boundaries of the time interval. For the sake of a later argument it is retained in Eq.~(I-6).
Eq.~(I-6) implies that $Re \,\dirint{\delta \Psi(t)}{ i \frac{\partial}{\partial t}-\hat{H}}{\Psi(t)} = 0$ (see Eq.~\ref{eq:vp5}), but mistakenly it is stated 
that `` the vanishing of the first two terms [in Eq.~I-6] is equivalent to
the td SE''. Here the possibility is disregarded that 
$ \dirint{\delta \Psi(t)}{ i \frac{\partial}{\partial t}-\hat{H}}{\Psi(t)}$ may vanish 
(for all allowed variations within a given parametrization of the wave function)
while $(i \frac{\partial}{\partial t}-\hat{H})\ket{\Psi(t)}$ does not.

Next the discussion focusses on the RG AIF and the corresponding VP for the td density 
(see Eqs.~\ref{eq:actint} and \ref{eq:vpn1}).  
It is stated that $\delta A[n] = 0$ readily implies Eq.~(I-9), reading
\begin{equation}
V(\bs{r},t) =  \frac{\delta A_0[n]}{\delta n(\bs{r},t)}
\end{equation}
Obviously, this is just a more explicit form of Eq.~(\ref{eq:vpn3}) obtained by
expressing $\delta A[n]$ in terms of a functional derivative with respect to the 
td density. To make contact between the two forms given by Eq.~(I-9) and Eq.~(\ref{eq:vpn3})
let us note that 
the specific (external) potential,
\begin{equation}
V[n] = \int d\bs{r} \int_{t_1}^{t_2} dt\, V(\bs{r},t) n(\bs{r},t)
\end{equation}
can be treated separately in a partitioning of the
$A[n]$ into a universal and a specific part, $A[n] = A_0[n] - V[n]$. This yields
\begin{equation}
\frac{\delta A[n]}{\delta n(\bs{r},t)} = \frac{\delta A_0[n]}{\delta n(\bs{r},t)} - V(\bs{r},t)
\end{equation}
However, as discussed above it has been tacitly assumed here
that the integrand in $A[n]$ at a given time $t$ depends only on $n(t)$.
If $\delta A[n]$ can be expressed  
in terms of functional derivatives at all, then one has to expect
that functional derivatives with respect to $\dot n(t)$, or even $\ddot n(t)$ will come into play.
So irrespective of the problem of the definiteness of $A[n]$, 
Eq.~(I-9) certainly is not appropriate.

While Eq.~(I-9) is seen as questionable, the
origin of the problem is located not in the AIF but rather in the VP, 
implying that the remedy should be a modification of the latter.
So let us see what the ``solution'' here is.

Performing the
algebra along Eqs.~(I-5,I-6) in the case of $\delta A[n] = 0$, 
the analogue to Eq.~(I-5) reads    
\begin{equation}
\label{eq:mV5}
\delta A[n] =   \int_{t_1}^{t_2} dt\,
\dirint{\delta \Psi[n,\delta n](t)}{i \frac{\partial}{\partial t}-\hat{H}}{\Psi[n](t)}
+ \int_{t_1}^{t_2} dt \, 
\dirint{\Psi[n](t)}{i \frac{\partial}{\partial t}-\hat{H}}{\delta \Psi[n,\delta n](t)} = 0
\end{equation}
where the variations of the td density vanish at the initial and final 
times, $\delta n(t_1) = \delta n(t_2) = 0$.
As mentioned above, Vignale rightly observes that  
the corresponding variation of the wave function may not vanish at the final time, that is,
\begin{equation}
\delta \Psi [n,\delta n](t_2) \neq 0
\end{equation}
Integration by parts then leads to to the result
\begin{equation}
\label{eq:mV6}
\delta A[n] =  2 \int_{t_1}^{t_2} dt\, Re \,
\dirint{\delta \Psi[n, \delta n](t)}{i \frac{\partial}{\partial t}-\hat{H}}
{\Psi[n](t)} + i \bracket{\Psi[n](t_2)}{\delta \Psi[n, \delta n](t_2)} = 0 
\end{equation}
where in contrast to the VP for the td wave function there is 
the possibly non-vanishing extra term 
$\bracket{\Psi[n](t_2)}{\delta \Psi[n, \delta n](t_2)}$ on the r.h.s..
Apparently, consistency with the exact solution of the td SE
can only be achieved if the first term on the r.h.s. of Eq.~(\ref{eq:mV6}) vanishes.
This seems to imply that the original VP is not valid and must be replaced 
by  
\begin{equation}
\label{eq:mVP)}
\delta A[n] - i \bracket{\Psi[n](t_2)}{\delta \Psi[n, \delta n](t_2)} = 0 
\end{equation}
It appears though that this conclusion is somewhat premature.

What Eq.~(\ref{eq:mV6}) shows is that the AIF VP for the td density 
is not equivalent to a VP of the (Dirac-Frenkel) form
\begin{equation}
\label{eq:mV7}
 Re \, \dirint{\delta \Psi[n, \delta n](t)}{i \frac{\partial}{\partial t}-\hat{H}}
{\Psi[n](t)} = 0 
\end{equation}
(Moreover, one may not assume that
$i \delta \Psi[n, \delta n](t)$ is an allowed variation, so that 
there is no analogue of Eq.~(\ref{eq:mV7}) for the imaginary part.)
While this is an interesting finding, 
it does not yet discredit the 
original AIF VP. Let us again recall the idea underlying the AIF VP: 
to find a density trajectory $n(t)$ such that the associated wave function 
$\Psi[n](t)$ minimizes (makes stationary) the expectation value of the 
td SE deviation operator in a given time interval $[t_1,t_2]$, where the 
manifold of allowed variations of $n(t)$ is restricted by 
$\delta n(t_1) = \delta n(t_2) = 0$. There is no \emph{a priori} reason
why this should not be consistent with an exact solution of the td SE. Let 
$n(t)$ be the density trajectory associated with the exact solution 
of the td SE $\Psi[n](t)$ (for a given initial value $\Psi (t_1)$). Obviously, then
the first term on the r.h.s. of Eq.~(\ref{eq:mV6}) vanishes, but the 
disturbing 2nd term could vanish here as well, because for this particular
$n(t)$ all variations $\delta \Psi[n, \delta n](t_2)$ consistent with 
$\delta n(t_2) = 0$ might be orthogonal to $\Psi[n](t_2)$, so that
\begin{equation}
\label{eq:mV8)}
\bracket{\Psi[n](t_2)}{\delta \Psi[n, \delta n](t_2)} = 0  
\end{equation}
Of course, this is not a positive proof for the consistency of
the AIF VP with an exact solution of the td SE
(a valid proof would presuppose the definiteness of the AIF). 
However, it discloses
a highly plausible loophole 
that has been entirely disregarded in I. This means
that the conclusion concerning the necessity of 
a modified VP is not stringent.

One might argue that the extra term in Eq.~(\ref{eq:mVP)}) does not do harm because it will (possibly)
vanish anyway for the exact td density. However, in the case of 
a non-exact AIF it would distort the  
the solution from the optimal one. Moreover, with regard to deriving
an EOM the constant extra term is just a nuisance. It forces one to
introduce the functional derivative of a ($N$-electron) wave function,
\begin{equation}
\label{eq:mV9}
\Psi[n](\bs{r}_1,\dots,\bs{r}_N,t_2) = \int d\bs{r} \int dt\, 
\frac{\delta \Psi[n](\bs{r_1},\dots,\bs{r}_N,t_2)}{\delta n(\bs{r},t)}
\delta n(\bs{r},t)
\end{equation} 
The functional derivative in the integrand is an awesome entity indeed, even if needed only for performing the overlap integral
in Eq.~(I-12), which according I
represents the final EOM. Let us note that here, as above, 
the possible emergence of more general functional derivatives is ignored. 

To summarize, there are two major misconceptions in 
paper I. First, it is supposed (without discussion) that the
RG AIF can be expressed entirely in terms of functional derivatives with respect
to the density. While this goes back to the original RG paper~\cite{run84:997},
it is certainly inadequate. One has to expect that 
functional derivatives with respect to the first (and possibly higher) time derivatives
of the td density come into play. So any conclusions based on a too
narrow representation 
in terms of density functional derivatives will be flawed. 

Second, the conclusion that the original AIF 
is not consistent with the exact solution of the td SE is not compelling.
There may be a plausible way out of Vignale's dilemma,
so that there is 
no need for a modification of the original VP.
The extra term advocated in I not only is incompatible with the intuitive idea of the AIF but
leads, moreover, to a hopeless complication in the aspired derivation of an EOM for 
td denities.

\section{Concluding remarks}

The problems associated with the
use of the RG AIF have a simple and rather obvious reason, namely, 
that it is not well-defined in the first place. The indefiniteness with respect
to an arbitrary constant due to the undetermined td phase in the wave functions
does not disappear in the variational procedure. In fact, Vignale's observation
that the density induced variation of the wave function does not vanish at the
final time, makes the latter strikingly manifest. One has to face the conclusion
that the RG AIF is ill-defined and cannot be used as a means to derive
a density-based EOM. 

Given that fundamental deficiency, one will not expect anything meaningful to emerge from
merely modifying the VP, as proposed in I. Generally speaking, a logically stringent 
argumentation based on a false premise could be useful in laying bare the
defect of the premise. However, as we have shown, the arguments in paper I 
are themselves hardly rigorous and partly even misleading.

Accepting the fact that the RG AIF is incurably ill-defined, one is referred back 
to the direct mapping-based foundation of TDDFT, a supposedly valid alternative 
(see, for example, Marques and Gross~\cite{mar04:427}). However,
one may wonder how the essentially mathematical RG1 mapping theorem could suffice to 
establish a physical (and causal) EOM for the time-evolution 
of a quantum system (see Refs.~\cite{sch07:022513,sch08:056502}).
Unfortunately, there is no second founding paper (replacing the discredited
Ref.~\cite{run84:997}), in which the mapping-based route
to time-dependent Kohn-Sham equations is fully disclosed and, thus, could serve
as the basis for a critical analysis. A crucial ingredient of the mapping foundation,
namely, the fact that the
td Kohn-Sham equations and the exchange correlation potential-functional herewith introduced 
can only be established as a fixed-point iteration scheme
 - though one not backed by a valid VP~\cite{sch08:056502} -
still has the status of an unpublished insider lore.

\section*{Acknowledgements}  
The author thanks H.-D. Meyer for illuminative discussions 
concerning the time-dependent variational principles in quantum mechanics.


\end{document}